\documentclass[aps,showpacs,twocolumn,superscriptaddress,prl,floatfix]{revtex4}

\usepackage{amssymb,amsmath,graphicx}
\usepackage{bm}

\begin{document}

\bibliographystyle{apsrev}
\preprint{}

\title{Tuning the Kondo effect with a mechanically controllable break junction}

\author{J.~J.~Parks}
\affiliation{Laboratory of Atomic and Solid State Physics, Cornell
University, Ithaca, New York 14853}

\author{A.~R.~Champagne}
\affiliation{Laboratory of Atomic and Solid State Physics, Cornell
University, Ithaca, New York 14853}

\author{G.~R.~Hutchison}
\affiliation{Department of Chemistry and Chemical Biology, Cornell
University, Ithaca, New York 14853}

\author{S.~Flores-Torres}
\affiliation{Department of Chemistry and Chemical Biology, Cornell
University, Ithaca, New York 14853}

\author{H.~D.~Abru\~na}
\affiliation{Department of Chemistry and Chemical Biology, Cornell
University, Ithaca, New York 14853}

\author{D.~C.~Ralph}
\affiliation{Laboratory of Atomic and Solid State Physics, Cornell
University, Ithaca, New York 14853}

\date{\today}


\begin{abstract}
We study electron transport through C$_{60}$ molecules in the Kondo regime
using a mechanically controllable break junction.  By varying the electrode
spacing, we are able to change both the width and height of the Kondo
resonance, indicating modification of the Kondo temperature and the relative
strength of coupling to the two electrodes.  The linear conductance
as a function of $T/T_\textrm{K}$ agrees with the scaling function
expected for the spin-1/2 Kondo problem.  We are also able to tune finite-bias
Kondo features which appear at the energy of the first C$_{60}$ intracage
vibrational mode.
\end{abstract}

\maketitle

The Kondo effect is a many-body phenomenon that can arise from the coupling
between a localized spin and a sea of conduction electrons.  At low
temperature, a spin-singlet state may be formed from a localized spin-1/2
electron and the delocalized Fermi sea, leading to a correlated state 
reflected in transport as a zero-bias conductance anomaly \cite{KondoTheory}.
This feature has been observed in devices containing lithographically-defined
quantum dots \cite{Kondo2DEG}, carbon nanotubes \cite{KondoNT}, and single 
molecules \cite{KondoCornell,KondoHarvard, KondoC_60, KondoFerromagnetism}.
Other more exotic Kondo effects involving higher spin states 
\cite{IntegerSpinKondo,TwoStageKondo}, non-equilibrium effects 
\cite{KondoOutofEquilibrium,SingletTripletKastner, NonEquilibriumNT}, and 
orbital degeneracies \cite{OrbitalKondo} have also been observed.

While the Kondo effect has been intensely studied, previous experiments on
nanotubes and molecules have lacked control over the molecule-electrode 
coupling.  In this Letter, we address the influence of coupling on transport 
in the Kondo regime by tuning the Kondo effect in C$_{60}$ 
molecules with a mechanically controllable break junction
(MCBJ) \cite{MBJ}.  By varying the electrode spacing, we show that both
the Kondo temperature $T_\textrm{K}$ and the magnitude of the zero-bias
conductance signal associated with the Kondo resonance are modified.  
These changes allow a determination of how the motion modifies the 
relative coupling of the molecule to the two electrodes.
The normalized linear conductance exhibits scaling behavior as 
expected within the theory of the Anderson Model.  The same devices can also 
exhibit finite-bias inelastic Kondo features at an energy that corresponds to
the lowest-energy intracage vibrational mode of C$_{60}$. Changes in electrode 
spacing can tune the energy and amplitude of these signals, in ways that 
present a challenge to existing theory.

\begin{figure}[b]
	\label{Figure1}
\includegraphics[width=8.6cm]{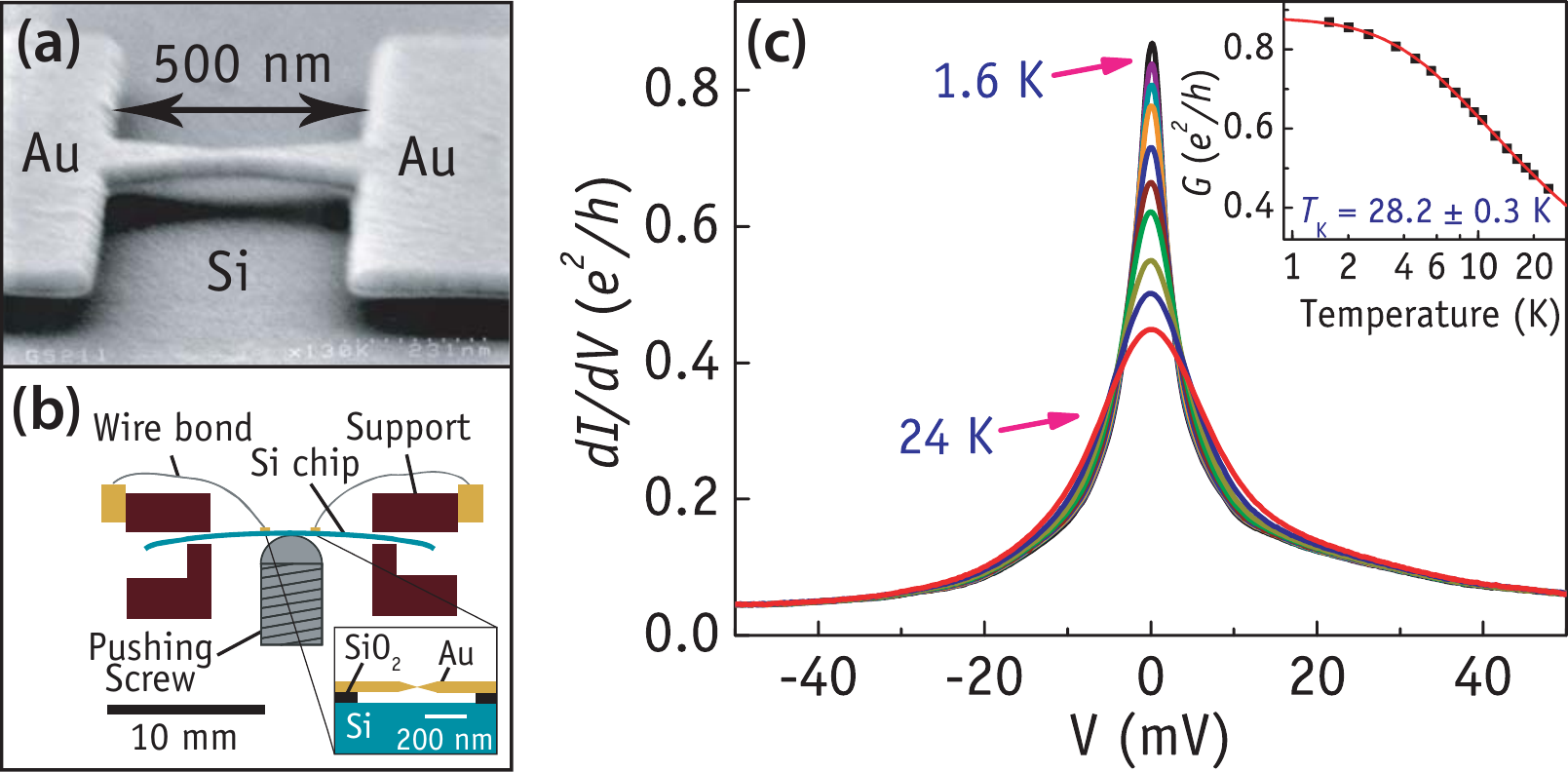} \caption{ \footnotesize{(color). 
(a) Scanning electron micrograph of a Au bridge suspended 40 nm above a
Si substrate. (b) Schematic of the MCBJ apparatus.
(c) $dI/dV$ traces of Device A at various temperatures. 
Inset: A fit of the linear conductance to the NRG interpolation function
(Eq. (\ref{NRG})) yields a Kondo temperature of 28.2 $\pm$ 0.3 K.}}
\end{figure}

Our devices consist initially of continuous gold lines (32 nm thick, 
500 nm long, and with a minimum width of 50 nm) suspended 40 nm above a 
200-$\mu$m-thick Si wafer \cite{MBJGate} (Fig. 1(a)). To incorporate 
C$_{60}$ molecules, we clean unbroken wires in an oxygen plasma 
(1 min at 30 mtorr, 0.25 W/cm$^2$) to remove any organic contaminants. 
We then deposit 25~${\mu}$L of a 100~${\mu}$M solution of C$_{60}$ in toluene,
wait 1 minute and blow dry. We cool the samples to 1.6~K.  The range of 
motion permitted by bending the Si substrate is generally too small to break
the Au wires by mechanical motion alone, so we use electromigration to create
a molecular-scale break in the wires before beginning studies as a function
of mechanical motion \cite{Electromigration}.  We minimize the series 
resistance in our circuit during the electromigration process (the total 
series resistance is $\approx$6~$\Omega$) and ramp a voltage stepwise until 
the resistance of a Au wire begins to change with time at constant 
$V$.  This occurs consistently for $V \approx 650$ mV.
We then hold $V$ fixed until the wire breaks.  After electromigration, we
find that one or a few molecules can sometimes ($\approx$20\% of 194 devices
studied) be found bridging the gap between electrodes, as inferred from the
existence of a Coulomb blockade characteristic in the $I$-$V$ curve rather
than a linear characteristic typical of bare junctions 
\cite{KondoCornell, KondoC_60}.  In approximately 3/4 of devices exhibiting
Coulomb blockade, $dI/dV$ displayed a peak at $V=0$, which is a signature of
the Kondo effect.  We performed control experiments with more than 200 bare
Au junctions as well as 76 junctions treated with toluene but without 
C$_{60}$, and found a zero-bias feature in $<$2\% of these devices: 3 bare Au
junctions and 1 device exposed to toluene.

We choose to study C$_{60}$ molecules because they are sufficiently durable 
to survive high temperatures present during
electromigration \cite{Trouwborst} and because previous work on
single-molecule C$_{60}$ devices has observed the Kondo effect 
\cite{KondoC_60, KondoFerromagnetism}.  Photoemission studies of C$_{60}$ 
in contact with Au have shown that the molecule tends to gain an electron 
from the Au due to the molecule's high electronegativity 
\cite{C60Charge1, C60Charge2}, so that in our work C$_{60}$ is likely to 
often possess an unpaired spin in equilibrium, providing conditions needed
for the Kondo effect.

We can control the size of the inter-electrode gap in our devices by using a
stepper motor attached to a pushing screw to bend the silicon substrate at
cryogenic temperatures (Fig. 1(b)).  To calibrate the changes in electrode
spacing, $d$, we measure the conductance of bare Au junctions as a
function of motor turns and fit to the tunneling conductance expression
$G\propto\exp(-2d\sqrt{2m_{e}\phi}/\hbar)$, where $\phi = 5.1$ eV is
the work function of Au. The mean calibration over 14 bare junctions is 
$6.1\pm 0.4$ pm per motor turn; we apply this value to determine electrode 
displacements in identically-prepared devices containing C$_{60}$.

Figure 1(c) shows differential conductance curves of a C$_{60}$ device 
(Device A) at several temperatures.  We observe a prominent zero-bias peak 
that is suppressed with increasing temperature, in accordance with 
predictions for the Kondo effect.  Shown in the inset of figure 1(c), we fit 
the linear response conductance as a function of temperature to an 
interpolation expression \cite{NRGexpt} that is a good fit to the numerical
renormalization group (NRG) result \cite{NRGTheory} for conductance in the 
Kondo regime with the addition of a constant background conductance
\cite{Pustilnik},
\begin{equation}
\label{NRG}
	G(T) = G_{0} \left[1+ \frac{T^2}{T_\textrm{K}^{2}}(2^{1/s} -
1) \right]^{-s} + G_{el},
\end{equation}
using  $s = 0.22$, and leaving $T_\textrm{K}$, $G_{0}$, and $G_{el}$ as free
parameters.  Our data are well described by this expression, and we extract
$T_\textrm{K} = 28.2 \pm 0.3$ K.
The value for the Kondo temperature also agrees with that obtained by setting
the full-width at half maximum (FWHM) of the base temperature zero-bias peak
(5.14~$\pm$~0.06~mV) to $2 k_{B} T_\textrm{K}/e$ \cite{KondoNT}, yielding
$T_\textrm{K} \approx 30$ K.

\begin{figure} \label{Figure2}
\includegraphics[width=8.6cm]{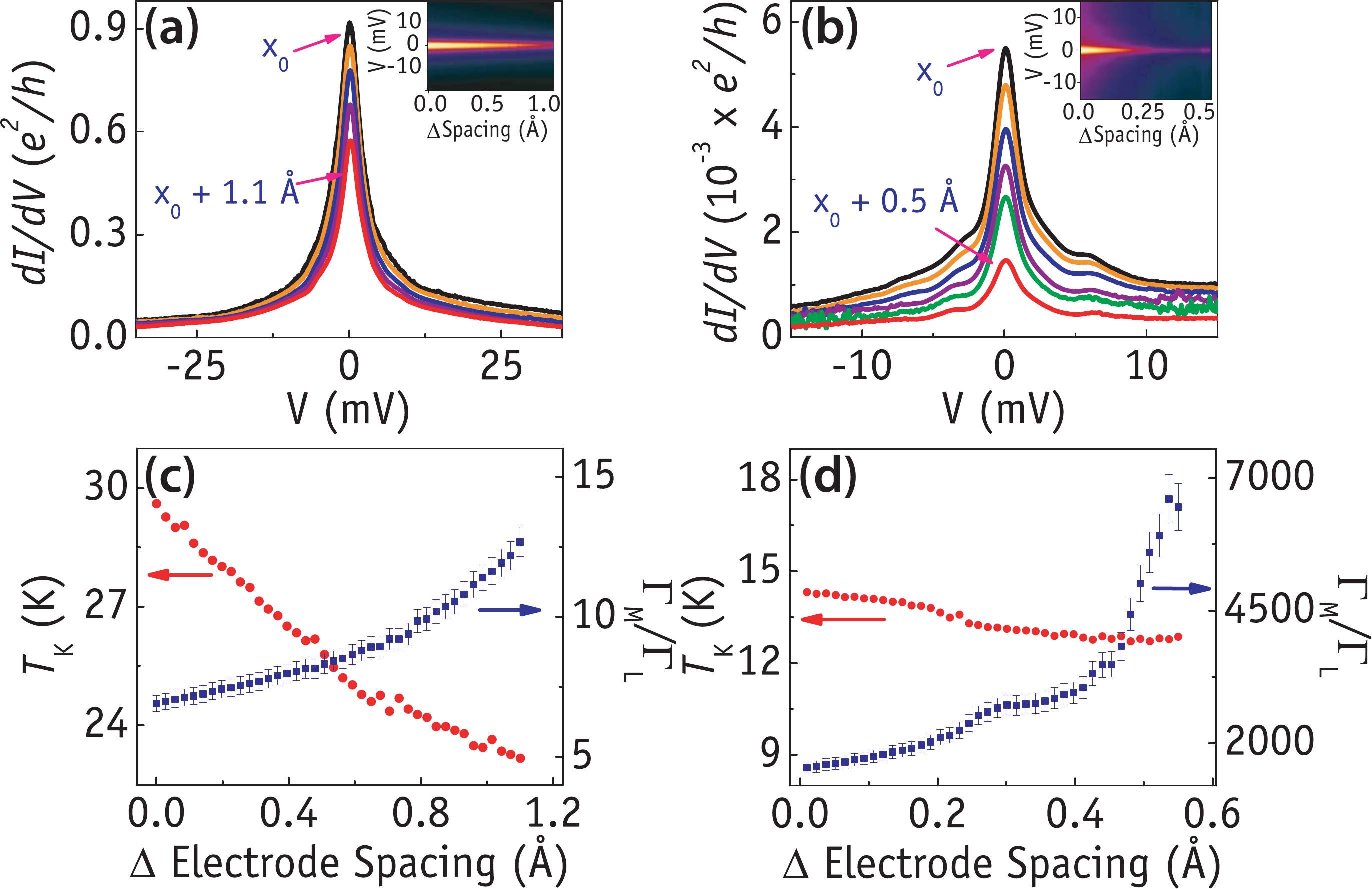} \caption{ \footnotesize{(color). 
$dI/dV$ traces of (a) Device A and (b) Device B at various
electrode spacings for $T =1.6$~K. Insets to (a) and (b): $dI/dV$ (in
color scale) as a function of bias voltage and electrode spacing at 
$T =1.6$~K. (c,d) The Kondo temperature and the relative coupling 
$\Gamma_{\textrm{M}}/\Gamma_{\textrm{L}}$ as a function of electrode spacing
for Devices A and B at $T =1.6$~K. The uncertainty in the determination of the
Kondo temperatures is smaller  than $\pm$0.4 K.}}
\end{figure}

Figure 2 displays the evolution of the zero-bias peak as we vary the electrode
spacing for Device A (Fig. 2(a)) and for Device B (Fig. 2(b)).  For each 
electrode spacing, we can determine the Kondo temperature from the FWHM
of the zero-bias resonance, and we can also deduce the relative coupling 
${\Gamma_\textrm{M}}/{\Gamma_\textrm{L}}$ of the molecule to the more- and 
less-strongly-coupled electrode based on the magnitude of the linear 
conductance near $T=0$ \cite{KondoTheory}
\begin{equation}
\label{linearconductance}
	G = \frac{2e^2}{h}  \frac{4\, \Gamma_\textrm{M}
\Gamma_\textrm{L}}{(\Gamma_\textrm{M} + \Gamma_\textrm{L})^2}
f(T/T_\textrm{K}) + G_{el}.
\end{equation}
To describe the dependence on $T$, we adopt Eq. (\ref{NRG}) and use
$f(T/T_\textrm{K}) = \left[1 + T^2/T_\textrm{K}^{2}(2^{1/s} - 1)\right]^{-s}$
with $s=0.22$.  The value of the background conductance is determined from
fits of the $dI/dV$ as a function of $V$ to a Lorentzian plus a
constant.  The low value of the peak conductance of Device B indicates
that the molecule is coupled quite asymmetrically to its electrodes.
In figures 2(c) and 2(d), we plot the evolution of the Kondo temperature and
relative coupling as a function of electrode spacing for Devices A and B.

We find that as the electrode spacing is varied,
${\Gamma_\textrm{M}}/{\Gamma_\textrm{L}}$ increases by $\approx$330\% for a 
displacement of 0.55 {\AA} in the asymmetrically-coupled Device B, while
the increase is just $\approx$80\% over a larger displacement of 1.1 {\AA} in
the more-symmetrically coupled Device A.  This suggests that as the electrodes
are pulled apart, the molecule in Device B remains well-coupled to one of the
leads so that the motion affects primarily $\Gamma_\textrm{L}$, whereas
in Device A, $\Gamma_\textrm{M}$ and $\Gamma_\textrm{L}$ are both modified,
although not exactly equally. The background conductance $G_{el}$
is always $\le 0.045$ $e^2/h$ in Device A and $\le 6.4 \times 10^{-4}$ $e^2/h$
in Device B, and decreases by approximately a factor of 10 with 1.1 {\AA} 
motion in Device A, and by approximately a factor of 3 over 0.55 {\AA} motion
in Device B.

In the more-symmetric Device A, the Kondo temperature can be tuned from
$\approx$30 K to $\approx$23 K as the inter-electrode spacing is increased,
whereas in asymmetric Device B, the Kondo temperature remains within a narrow
range of 13-14~K.  We can analyze these changes using the Haldane expression
for the Kondo temperature in the limit of large charging energy $U$ 
\cite{Haldane}, 
\begin{equation}
\label{kondotemp}
	T_\textrm{K} = \frac{\sqrt{\Gamma U}}{2}e^{\pi
\varepsilon_{0}(\varepsilon_{0}+U)/\Gamma U} \sim e^{\pi
\varepsilon_{0}/\Gamma}
\end{equation}
where $\Gamma = \Gamma_\textrm{M} + \Gamma_\textrm{L}$ and $\varepsilon_{0}$
is the energy relative to the Fermi level of the localized state that produces
the Kondo effect. We can expect both
$\Gamma$ and $\varepsilon_{0}$ to vary as a function of electrode
spacing: $\Gamma$ because the
coupling of the molecule to at least one of the electrodes must decrease as the
electrodes are moved apart and $\varepsilon_{0}$ because break junctions
generally exhibit large built-in electric fields even when $V$=0, so that
motion of the electrodes produces a gating effect on energy levels in the
molecule \cite{MBJGate}.  If $\Gamma_\textrm{M}$ and $\Gamma_\textrm{L}$ are
significantly asymmetric, then varying the electrode spacing will likely have
little effect on the overall $\Gamma$, as only the weaker coupling may
change significantly.  This regime applies to Device B, where the coupling
ratio always exceeds 1500.  The observation of only a small change in
$T_\textrm{K}$ as a function of electrode displacement for Device B is
consistent with this picture. For Device A, we cannot distinguish the relative
contributions of changes in $\Gamma$ and $\varepsilon_{0}$ to the tuning of
$T_\textrm{K}$, based on our data. In principle, a gate electrode that can
independently adjust $\varepsilon_{0}$ could help to disentangle the effects
of adjusting electrode spacing.  However, we find that the gate coupling
for our device geometry is too weak to adjust $\varepsilon_{0}$ measurably for
devices in the Kondo regime.

\begin{figure}[t]
       \label{Figure3}
       \includegraphics[width=8.6cm]{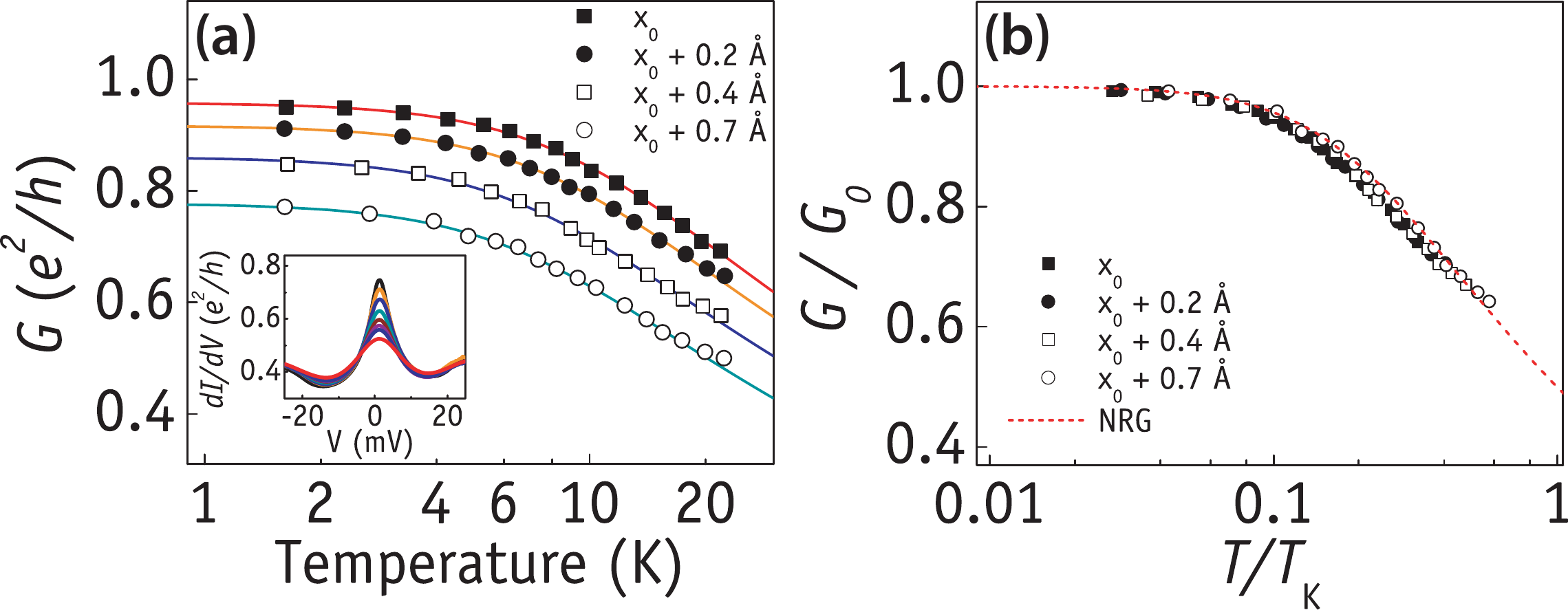}
\caption{ \footnotesize{(color). 
(a) Linear conductance of Device C as a function of temperature, fit to the 
NRG interpolation expression (Eq. (\ref{NRG})).  The extracted Kondo 
temperatures are (from the top trace to the bottom): $60.3\pm2.4$~K,
$55.5 \pm 0.9$~K, $45.6 \pm 1.9$~K, $38.1 \pm 1.2$~K.  Inset:
$dI/dV$ traces at $x_{0} + 0.7$ \AA. (b) The normalized conductance,
$G(T/T_\textrm{K})/G_0$, is a universal function of $T/T_\textrm{K}$.
Scaled conductance data is compared to the NRG calculation (dotted line).}}
\end{figure}

In figure 3(a), we show the temperature dependence of linear conductance for
Device C at several different electrode spacings.  In this device, the Kondo
temperature could be tuned over a significant range, from $60.3 \pm
2.4$ K (top curve)
to $38.1 \pm 1.2$ K (bottom curve). In the Kondo regime, the conductance
normalized to its zero-temperature value is expected to be dependent solely on
$T/T_\textrm{K}$ and thus to exhibit universal scaling behavior.
In figure 3(b), we show that $G(T/T_\textrm{K})/G_{0}$ does indeed exhibit a
reasonable collapse onto a function that is in close agreement with
the universal
function predicted by NRG calculations \cite{NRGTheory}.

The evolution of transport properties as a function of electrode spacing is 
not always as smooth as we measured for Devices A, B, and C.  In other
devices, the conductance could exhibit sudden changes, and zero-bias 
conductance resonances could fluctuate in and out of existence.  We ascribe
these abrupt changes to motion of the C$_{60}$ molecule within the junction
region.

In addition to a zero-bias peak in $dI/dV$, in 5 out of 23 devices with Kondo
temperatures greater than 20 K we have also observed peaks in $dI/dV$ at 
symmetric values of $V$ near $\pm$33 mV (specifically, at 29.6, 32.8, 33.5,
36.9, and 37.2 mV in the five devices); see figure 4(a).  We did not observe
any other similar features for $|V| < 60$ mV. The energy of 33 meV is known to
correspond to the lowest intracage vibrational mode of isolated C$_{60}$, in 
which the molecule oscillates between a sphere and a prolate ellipsoid shape 
(Fig. 4(a), inset) \cite{C60Vibrations}.  Previous investigations have shown 
that molecular vibrations can enhance $dI/dV$ at energies corresponding to 
vibrational quanta \cite{C60, vanRuitenbeekstretching, C140}. For devices in 
the Kondo regime, coupling to a vibrational mode has been predicted to result
in an inelastic Kondo effect, which is manifested as sidebands in $dI/dV$ at 
$V\neq0$ \cite{KondoVibrations}.  Finite-bias features in
the Kondo regime have been observed in single-molecule transistors coupled to
a vibrational mode \cite{KondoCornell, KondoC_60, NatelsonInelastic}, in 
quantum dots coupled to an applied microwave field \cite{KondoMicrowave}, 
and in quantum dots due to Kondo screening of excited states 
\cite{SingletTripletKastner,NonEquilibriumNT,MultipeakKondo}.

Figures 4(b) and 4(c) show $d^2I/dV^2$ for Devices D and E as a function
of bias voltage and electrode spacing. Consistent with what we found for
Devices A and B, as the electrodes are pulled apart the magnitude of the
zero-bias peak decreases more strongly in the less-symmetrically coupled
(lower conductance) Device E than in the more-symmetrically coupled Device D.
The strength of the sidebands is also different in the two devices; the
satellite peaks are significantly more prominent in the more-symmetrically
coupled Device D  compared to the less-symmetrically coupled Device E, as
predicted in Ref. \cite{KondoVibrations}.  As the electrode spacing, and hence
the coupling asymmetry, is increased, the amplitude of the non-equilibrium
peaks decreases in Device D, but the small peaks of Device E do not seem to
be strongly modified. In both devices, the positions of the inelastic features
increase in $|V|$ as the electrodes are pulled apart, suggesting that the
mechanical motion increases the energy of the active vibrational mode.

The changes in vibrational frequency as a function of mechanical motion are
larger than what we anticipated based on molecular modeling.  We performed 
calculations in Gaussian 03 using the PM3 semiempirical Hamiltonian, which
has been accurate in predicting the vibrational frequencies of fullerenes
\cite{C60PM3}.  The geometries and harmonic frequencies were calculated under
$C_{2h}$ symmetry using the native structure and by setting distance 
constraints on two atoms at opposite ends of the neutrally charged C$_{60}$ 
cage to define the long axis from the native length of 7.092~{\AA} to 
7.792~{\AA}.  The calculations indicate that the five-fold degenerate 
$H_{g}(1)$ mode at 33 meV is broken upon distortion into a set of two 
nearly-degenerate ``short-axis'' modes (involving motion perpendicular to the 
direction of stretching) which decrease in energy as the molecule is stretched 
and three ``long-axis'' modes which increase in energy (Fig. 4(d)). Only 
increases in energy are observed experimentally; we speculate that the 
long-axis modes may couple more strongly to electron transport. If we assume 
that the increase in molecular diameter is equal to the increase in the 
electrode spacing, so as to determine an upper limit for the estimated 
frequency shift, then we find that the measured shift for Device E is 
comparable to the calculated frequencies of the long-axis modes, with the main
deviations coming at large molecular diameter.  However, the peak positions in
the more-symmetrically coupled Device D shift more strongly than predicted, by
roughly a factor of 10.  This discrepancy suggests that more rigorous 
theoretical work may be needed to understand electron-vibration coupling in
single-molecule systems, when including coupling to the electrodes and the 
Kondo effect.

\begin{figure}[t]
       \label{Figure4}
       \includegraphics[width=8.6cm]{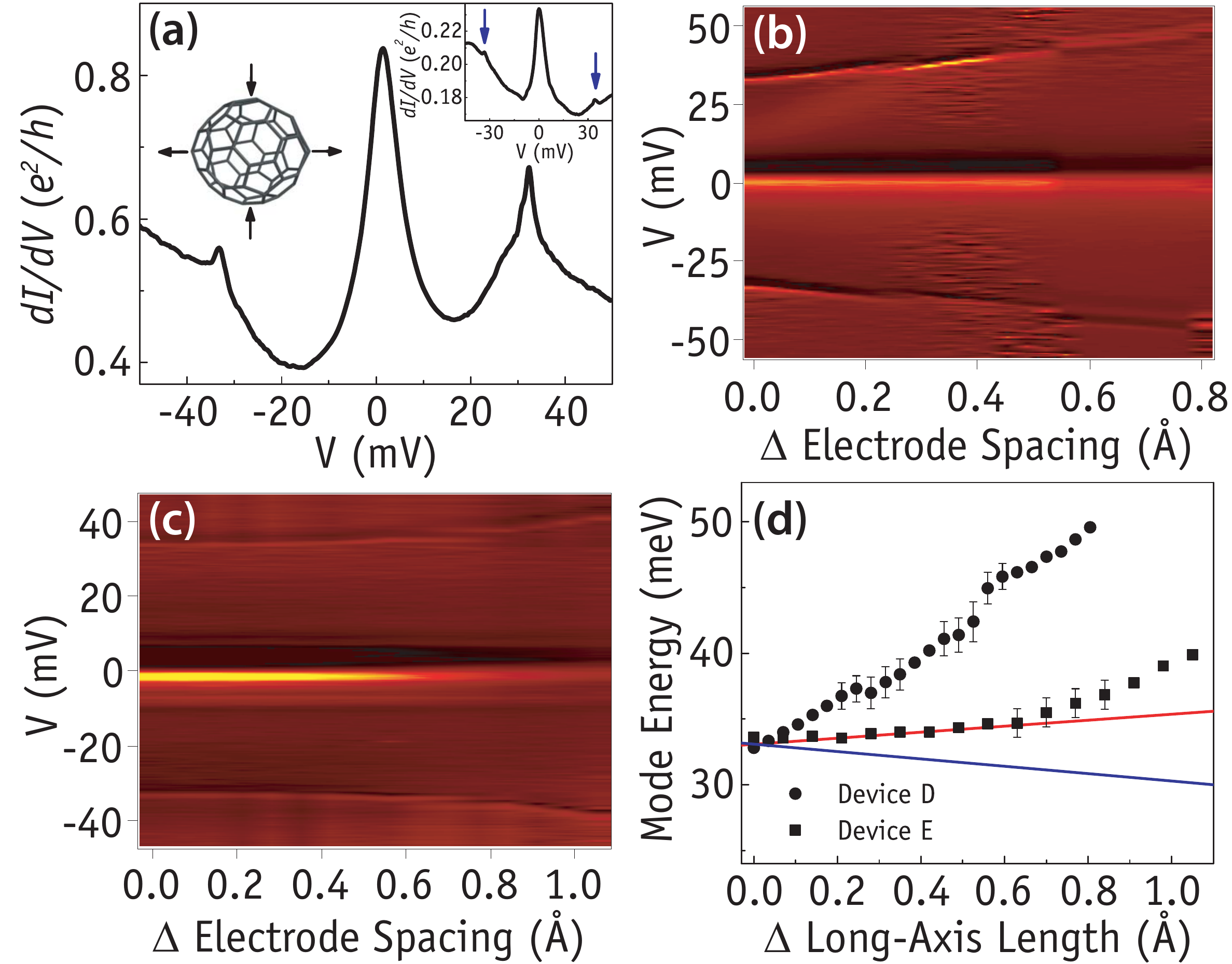}
\caption{ \footnotesize{(color). 
(a) $dI/dV$ for Device D at $T =1.6$~K showing
satellite peaks near $\pm$33 mV. Left inset: Schematic of the
$H_{g}(1)$ intracage vibrational mode.
Right inset: $dI/dV$ for Device E at $T =1.6$~K.
(b, c) $d^2I/dV^2$ as a function of bias voltage and electrode spacing for
Devices D and E at $T$~=~1.6~K. (d) Stretching dependence of mode energies 
from Devices D and E.  Lines: short-axis (negative slope) and long-axis
(positive slope)
vibrational energies calculated using the PM3 semiempirical method.}}
\end{figure}

In summary, we have demonstrated how the Kondo effect is modified by
tuning the spacing between electrodes in mechanically controllable break
junction devices containing C$_{60}$ molecules. We measure changes in both
the Kondo temperature and zero-bias conductance that are in good accord with 
theoretical expectations and that allow us to characterize how the 
mechanical motion changes the relative coupling of the molecule to the 
electrodes.  We have also observed and tuned finite-bias Kondo features which
appear at energies corresponding to an intracage vibrational mode of C$_{60}$.
We find that the vibrational energy can change more strongly as a function of 
stretching than predicted by a simple semiempirical Hamiltonian, thereby 
presenting a challenge for more accurate theory.

We thank D. Goldhaber-Gordon, P. Brouwer, and A. Pasupathy for discussions, 
and K. Bolotin, F. Kuemmeth, and J. Grose for discussions and experimental 
help.  JJP thanks the NSF for graduate fellowship support.  This work was
supported by the NSF (DMR-0244713, CHE-0403806, DMR-0605742), the NSF/NSEC
program through the Cornell Center for Nanoscale Systems, and through use of 
the Cornell Nanoscale Facility/NNIN.

\end{document}